\documentclass[conference]{IEEEtran}
\IEEEoverridecommandlockouts
\usepackage{cite}
\usepackage{amsmath,amssymb,amsfonts}
\usepackage{algorithmic}
\usepackage{graphicx}
\usepackage{textcomp}
\usepackage{xcolor}

\usepackage{color, soul}
\usepackage{multirow}
\graphicspath{{./fig/}}
\DeclareGraphicsExtensions{.pdf,.jpeg,.png}
\def\BibTeX{{\rm B\kern-.05em{\sc i\kern-.025em b}\kern-.08em
    T\kern-.1667em\lower.7ex\hbox{E}\kern-.125emX}}
\begin{document}

\title{A Linux Kernel Scheduler Extension for Multi-Core Systems}

\author{\IEEEauthorblockN{Aleix Roca\IEEEauthorrefmark{1},
Samuel Rodriguez\IEEEauthorrefmark{1}, 
Albert Segura\IEEEauthorrefmark{1},
Kevin Marquet\IEEEauthorrefmark{2} and
Vicen\c{c} Beltran\IEEEauthorrefmark{1}}

\IEEEauthorblockA{\IEEEauthorrefmark{1}Barcelona Supercomputing Center, Jordi Girona 29-31, Barcelona, 08034, Spain\\
Email:\{arocanon, samuel.rodriguez, asegurasalvador, vbeltran\}@bsc.es}
\IEEEauthorblockA{\IEEEauthorrefmark{2}Univ Lyon, INSA Lyon, Inria, CITI, 6 avenue des arts, Villeurbanne, F-69621, France\\
Email: Kevin.Marquet@insa-lyon.fr}}

\maketitle

\begin{abstract}
%Linux kernel is used on many environments, but it is mostly designed to run multi-programed environments. High-performance applications has special characteristics (no multi-programming)
% The core of a blocked thread => 
	
The Linux kernel is mostly designed for multi-programed environments, but high-performance applications have other requirements. Such applications are run standalone, and usually rely on runtime systems to distribute the application's workload on worker threads, one per core. However, due to current OSes limitations, it is not feasible to track whether workers are actually running or blocked due to, for instance, a requested resource. For I/O intensive applications, this leads to a significant performance degradation given that the core of a blocked thread becomes idle until it is able to run again. In this paper, we present the proof-of-concept of a Linux kernel extension denoted \textit{User-Monitored Threads} (UMT) which tackles this problem. Our extension allows a user-space process to be notified of when the selected threads become blocked or unblocked, making it possible for a runtime to schedule additional work on the idle core. We implemented the extension on the Linux Kernel 5.1 and adapted the \textit{Nanos6} runtime of the OmpSs-2 programming model to take advantage of it. The whole prototype was tested on two applications which, on the tested hardware and the appropriate conditions, reported speedups of almost 2x.

\end{abstract}


\section{Introduction}

High-performance computing applications usually rely on \textit{task-based programming models} to parallelize and seamlessly load balance an application's workload. The main objective of a programming model is to provide an abstraction layer for application developers that eases the task of getting the most out of available hardware resources. For this purpose, task-based programming models rely on \textit{runtime systems} to schedule fragments of an application's work (named tasks) in threads (named workers) as well as to manage the tasks execution sequence.

Runtime systems, due to its particular nature, usually perform its own HPC-tailored thread scheduling on top of the \textit{Operating System} (OS) scheduler. The objective is to maximize data cache reusability and memory locality in NUMA machines\footnote{Runtimes, unlike the kernel, know in advance which data a task will access as specified on its dependencies. Therefore, it is better for a runtime to distribute tasks on pinned workers than let the kernel guess where each worker should run based on its previous accesses.} given that runtimes have a more in-depth knowledge of the application's behaviour than the general OS scheduler.

The usual pattern followed by runtimes is to bound a worker per available core\footnote{Throughout this paper, we use the therm "core" to designate a logical computation unit such as a hardware thread.} with the objective of minimizing thread migrations and oversubscription, being both sources of cache pollution. Oversubscription refers to a period of time in which multiple threads are competing for the same core while in the ready state. This is generally undesirable given the OS' context switch overhead and the penalization incurred for having to share the core's caches.

However, runtime's balancing capabilities are subject to the underlying OS scheduler policy. In particular, when a thread performs a blocking I/O operation against the OS kernel, the core where the thread was running becomes idle until the operation completes. Certainly, although runtimes only keep a worker bound per core to avoid oversubscription, other non-runtime threads such as kernel or system background threads might be scheduled in the meantime. Nonetheless, because I/O operations are generally expensive, most of the time the core is likely to remain idle. This problem can lead to significant performance loss as some HPC or high-end server applications perform lots of I/O operations while dealing with file and network requests.

A possible solution is to make the runtime aware of blocked and unblocked workers. In this way, a blocked worker's core could be used by another worker in the meantime. This approach requires special kernel support, and several solutions exist to do so, but their complexity has prevented them from inclusion into the Linux kernel mainline code.

In this article, we make the following contributions:
\begin{itemize}
\item We propose a new, simple and lightweight Linux kernel extension in which the OS provides user-space feedback on the blocking and unblocking activities of its threads.
\item We modify the \textit{Nanos6} runtime of the \textit{OmpSs-2}~\cite{7967171} task-based programming model to take advantage of the proposed Linux Kernel extension.
\item We evaluate the whole prototype experimentally and show that speedup of up to 2x can be achieved.
\end{itemize}

Although the focus of this article is on the integration of our solution with runtime systems, it is worth noting that its scope is wider, being usable on other applications that both rely on I/O and a multi-threaded environment.

% removed because of lack of space
%The remainder of the paper is organized as follows. Section~\ref{sec:related-works} presents the related works. Section~\ref{sec:proposal} details our proposal, then Section~\ref{sec:experiments} gives experimental validation results. Finally, Section~\ref{sec:conclusion} concludes and gives some perspectives on this work.

\section{Related Work}
\label{sec:related-works}

%TODO: look at this new paper from vic,ens https://arxiv.org/abs/1906.08239

Mechanisms to provide kernel-space to user-space feedback when a blocking operation occurs have already been considered in the context of \textit{user-level threads}. User-level threads provide means to perform context switches in user-space, thus minimizing the cost of context switching. This is also known as the N:1 model, in which N user-level threads are mapped to a single kernel thread. The problem with this model is that when a user-level thread performs a blocking operation, all user-level threads associated with the same kernel-level thread block. A palliative approach exists, known as the \textit{Hybrid approach} in which a set of user-level threads are mapped to a set of kernel threads. However, if just one of these user-level threads block, all the other user-level threads associated with the same kernel-level thread will block. The heart of the matter is that the kernel is not aware of user-level threads.

Scheduler Activations (SA)~\cite{anderson:SA} provides user-level threads with their own reusable kernel context and a user-to-kernel feedback mechanism based on \textit{upcalls} (function calls from kernel-space to user-space). When a user-space thread blocks, a new type of kernel thread known as \textit{activation thread}, is created (or retrieved from a pool) to relieve it. The activation thread upcalls a special user-space function that informs the user-space scheduler of the blocked thread. Then, still on the upcall, the user scheduler runs and schedules another user-space thread. When the blocking operation finishes, another activation thread upcalls an user-space function to inform the user scheduler that its thread is ready again. 

% SA paper, page 9, section 3.1
%
% [...]
%
% When a program is started, the kernel creates a scheduler activation,
% assigns it to a processor, and upcalls into the application address space at a
% fixed entry point. The user-level thread management system receives the
% upcall and uses that activation as the context in which to initialize itself and
% run the main application thread. As the first thread executes, it may create
% more user threads and request additional processors. In this case, the kernel
% will create an additional scheduler activation for each processor and use it to
% upcall into the user level to tell it that the new processor is available. The
% user level then selects and executes a thread in the context of that activation.
%
% Similarly, when the kernel needs to notify the user level of an event, the
% kernel creates a scheduler activation, assigns it to a processor, and upcalls
% into the application address space. Once the upcall is started, the activation
% is similar to a traditional kernel thread— it can be used to process the event,
% run user-level threads, and trap into and block within the kernel.
%
% [...]

SA has a significant drawback: the user-space scheduler thread cannot safely access shared resources protected with a lock that it has no direct access to. For example, consider a user-level thread blocked on a page fault while holding an internal glibc lock. In response, the SA kernel would wake up another user-level thread to handle the blocking event. If the user-level code event handler were to acquire the same glibc resource as the blocked thread, it would deadlock. This extends to internal kernel structures, such as memory allocation locks.

% KSE support removed 
%https://en.wikipedia.org/wiki/FreeBSD_version_history
%https://www.freebsd.org/releases/7.0R/relnotes.html
%https://lists.freebsd.org/pipermail/freebsd-current/2008-March/084248.html

SA was integrated into production OS's such as NetBSD \cite{williams:BSDSA1, Small:BSDSA2} and FreeBSD \cite{evans:FreeBSDSA} (known as Kernel Schedule Entities or KSE). A Linux implementation was proposed \cite{danjean:LinuxSA1, danjean:LinuxSA2, danjean:LinuxSA3} but the SA concept was rejected because of its complexity \cite{WEB:LinuxSA}. In the NetBSD 5.0 version, SA support was removed for the previous 1:1 threading scheme because \textit{"The SA implementation was complicated, scaled poorly on multiprocessor systems and had no support for real-time applications"}~\cite{rasiukevicius:BSDRemove}. FreeBSD KSE support was also dropped since version 7.0.

Windows OS has a similar implementation called User-Mode Scheduling (UMS)~\cite{windows:ums}, it is based on the same principle of upcalls, userland context switches and in-kernel unblocked thread retention. The interface is available since the 64bit version of Windows 7 and Windows Server 2008 R2. The locking problem also arises in their implementation as noted in the cited document above: \textit{"To help prevent deadlocks, the UMS scheduler thread should not share locks with UMS workers. This includes both application-created locks and system locks that are acquired indirectly by operations such as allocating from the heap or loading DLLs"}.

The K42 research OS \cite{appavoo:k42_sched,krieger:k42_build} proposed a more sophisticated mechanism to solve a similar problem. The K42 kernel schedules entities called dispatchers, and dispatchers schedule user-space threads. A process consists of an address space and one or more dispatchers. All threads belonging to a dispatcher are bound to the same core as the dispatcher is. Hence, to achieve parallelism, multiple dispatchers are required. 

When a user-space thread invokes a kernel service to initiate an I/O request, a "reserved" thread from the kernel space is dynamically assigned from a pool of reserved threads. This thread is in charge of initiating the I/O operation against the underlying hardware and block until the request is ready. In the meantime, the kernel returns control to the user-space thread dispatcher so it can schedule another user thread. When the I/O completes, the kernel notifies the dispatcher with a signal-like mechanism so it can schedule the user thread again. It is worth noting that because dispatchers schedule user threads, an unblocked thread is not going to run unless there is some explicit interaction from the dispatcher scheduler.

The K42 user-level dispatcher's scheduler is provided by a trusted thread library. However, this library suffers from the same problem as SA: it cannot share any lock with the user-level threads; otherwise, a deadlock would block the process if the dispatcher's scheduler tried to get a lock that was already taken by the blocked user thread.

The proposed \textit{User-Monitored Thread} (UMT) is similar to SA and K42 in the sense that both use a mechanism to notify a user-space thread whenever another thread blocks or unblocks in kernel-space. The main differences of UMT with SA and K42 are, on the one hand, that unblocked threads are not retained anywhere (hence, a deadlock with the user-space scheduler is not possible) and, on the other hand, the UMT implementation is much more simple and lightweight. However, as a consequence of not retaining unblocked threads, UMT needs to deal with periods of oversubscription. Nonetheless, the UMT correction mechanisms minimize its effect.

Essentially, APIs than asynchronously "notify" a user program with an upcall rather than "return" information with a downcall, are more complex; similarly to how signal handlers compare to the signalfd mechanism. UMT simplicity partially lies on its downcall-based approach.

TAMPI~\cite{bsc:tampi} and TASIO~\cite{bsc:tasio} libraries tackle a similar problem with a completely different approach. Both libraries rely on the OmpSs-2 asynchronous-aware feature to enable I/O and computation overlapping by integrating asynchronous operations into the tasking model. TAMPI works at the MPI layer while TASIO at the OS surface. In essence, they translate synchronous operations to asynchronous and, instead of blocking, they return control back to the runtime. In general, TAMPI and TASIO are ad-hoc solutions that require blocking and non-blocking APIs while UMT is a generic approach that works with any blocking event regardless of existing asynchronous support. However, because of their specialization, the ad-hoc solution performance could be better in some situations.

\section{Proposal: User-Monitored Threads (UMT)}
\label{sec:proposal}                                                             

% removed because of space limitations
%In this Section we detail the UMT proposal and prototype implementation. First, we give a general overview and, next, we detail the main two UMT components: the kernel-space support and the user-space runtime coupling.

\subsection{Proposal overview}
\label{sec:prop-overview}

\begin{figure}
\includegraphics[width=\columnwidth]{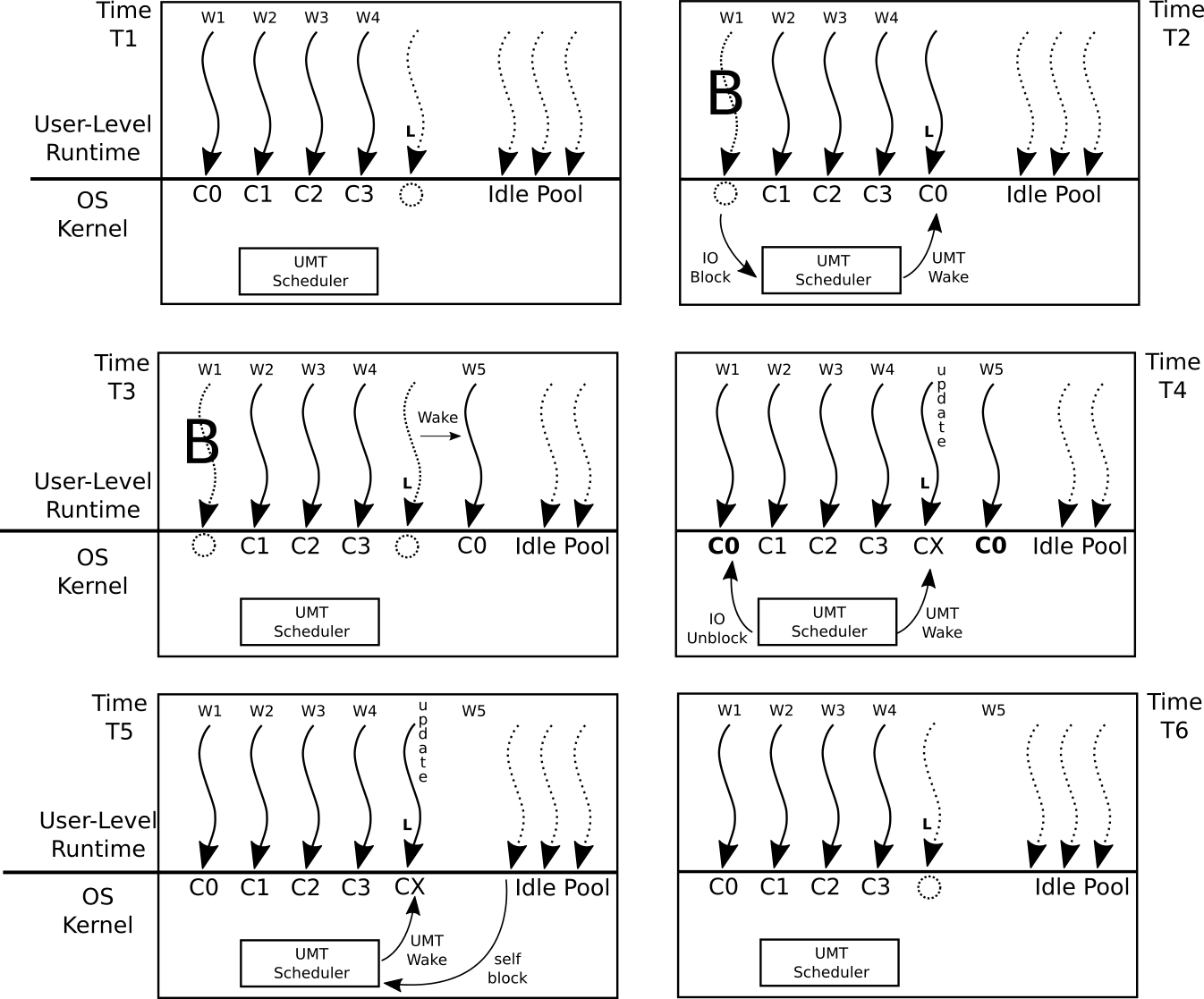}
\caption{UMT overview example}
\label{fig:overview}
\end{figure}

The \textit{User-Monitored Threads} (UMT) model specifies a mechanism that allows user applications to receive Linux kernel notification on the blocking/unblocking events of a subset of its threads. The details of this functionality are given in the following sections, but they are sketched in Fig.~\ref{fig:overview}. In this figure, the \textit{UMT Scheduler} box illustrates the modified Linux Kernel process scheduler, $W_i$ identifies runtime's worker threads, and $L$ denotes the runtime's \textit{Leader Thread} whose role is to monitor the UMT communication channel. $L$ is free to run in any core; the OS scheduler decides which worker will be preempted for L to run if any. Basically:

\begin{itemize}
\item At time T1, four workers W1, W2, W3 and W4 are bound to cores C0, C1, C2 and C3 respectively. L is waiting for UMT events. An idle pool of blocked workers holds.

\item At time T2, the worker W1 blocks because of an I/O operation and L is notified of the event.

\item At time T3, L wakes an idle worker from the pool and waits again for more events (when W5 wakes, it would also generate an unblock event which is omitted for simplicity). Worker W5 is now running on a core; without the proposed mechanism, it would have been idle.

\item At time T4, W1 is unblocked after the I/O operation finishes. An unblocking event is generated and L wakes up. Because there are no free cores at the moment, L waits until it momentously preempts another worker. Once it does so, it reads the UMT events and notices that multiple workers (W1 and W5) are running on the same core (C0).

\item At time T5, after W5 finishes executing tasks, it checks L's current UMT events status and realizes that there is an oversubscription problem affecting its current core. To fix the problem, the worker self surrenders and returns to the idle pool. This generates another event that wakes up L, which updates the UMT events status again.

\item At time T6, the oversubscription period has ended and the four workers are running normally.

\end{itemize}

\subsection{Kernel-space support}
\label{sec:notif-chann-betw}

The UMT kernel-space support relies on two components: A pair of new system calls, which are used to initiate and manage UMT, and the \textit{eventfd} Linux kernel feature, used as the notification channel between kernel- and user-space.

An eventfd is a simplified pipe designed as a lightweight inter-process synchronization mechanism. Eventfds are interfaced as usual file descriptors but, internally, they simply hold a 64 bit counter. The standard \texttt{write()} and \texttt{read()} system calls can be used to increment and read the counter, respectively. Once read, the counter is cleared, but if its value was zero, the reader blocks until something is written.

In UMT, eventfds are created when a user-space process claims that it wants to monitor some of its threads by calling the new system call \textit{umt\_enable()}. At this point, the Linux Kernel initializes an eventfd per core, stores them in the context of the calling process and returns them to user-space. Threads start being monitored as soon as each of them call the \textit{umt\_thread\_ctrl()} syscall. UMT uses each eventfd to simultaneously keep the count of both blocked and unblocked threads~\footnote{UMT does not keep track of preempted threads. Whenever a thread is preempted, its core does not become idle, but another thread starts running on it. Therefore, it is not necessary to inform user-space on such event.} in the corresponding core since the last read. The block and unblock counters are stored in the first 32 and the next 32 bits of the eventfd's 64 bit counter, respectively. Counter overflows\footnote{Namely, if $2^{32}$ monitored threads were to block on the same core without the core's eventfd being read, the blocked counter would overflow and unnoticeably corrupt the unblocked counter.} are not considered as we decided to focus on simplicity and UMT viability first. Future UMT versions will not rely on these counters, as mentioned in Section \ref{sec:discussion}.

Within the kernel, the eventfds are written exclusively in a new wrapper function that substitutes the genuine Linux Kernel \texttt{\_\_schedule()} function. This function is the common entry point of all paths that leads to a context switch. The blocked eventfd counter is incremented by one just before calling the genuine \texttt{\_\_schedule()} function while the unblocked eventfd counter is incremented by one on return. Not all threads that call \texttt{\_\_schedule()} do block; some of them might just be preempted. Only threads about to block or that wake up after being blocked (regardless of the reason) do write the eventfds counters. The right scenario is easily detected by just checking the current process' internal Linux state to be equal to the TASK\_RUNNING macro in the case of blocking and by checking the previous running state in the case of unblocking. 

The eventfd counters are consumed by a user-space application through the standard \texttt{read()} system call, which resets the eventfd counters to zero. The eventfd read value holds the number of blocked and unblocked tasks on the core since the last read operation. By subtracting the number of blocked threads to the number of unblocked threads, the number of ready threads in the associated core's eventfd is known. However, because each read operation erases the count, it is necessary to keep a user-space per core count and add the result of the subtraction after each eventfd read.

A thread migration might lead to uncompensated counters i.e. a thread migrated from core A to core B while it was in the preempted state in core A, will not have the chance to trigger the eventfd block event on core A before migrating and, therefore, it will be seen as if the thread is still running in A even after being migrated to B. When a process migration occurs, the UMT patch ensures the consistency of eventfd counters by checking, after a context switch, if the current thread core differs from its last one. If this is the case, a block event might have been missed on the previous core and, if needed, the missed event is written on the previous core's eventfd at this point (see Section \ref{sec:discussion} for more details).

Because of UMT's kernel-side simplicity, and unlike other similar approaches, it is likely that this code will not conflict with other Linux Kernel features and be easy to maintain. Also, the introduced overhead for non-UMT-enabled applications is kept to a minimum given that the UMT instrumentation points are reduced to two new conditional statements in the context switch path. However, the interface used in this implementation could evolve in future versions to match the Linux Kernel community requirements. 

\subsection{User-space support: The case of Nanos6}
\label{sec:userl-runt-syst}

The UMT user-space support requires four main components: An initialization step to enable the UMT kernel feature (using the new system calls), a mechanism to read and process the eventfds counters of each core (using the \textit{read()} syscall) to obtain the number of ready threads, a mechanism to wake up threads on idle cores based on the counter of ready threads, and a protection mechanism to limit oversubscription.

In order to validate the proposal, we have adapted the \textit{Nanos6} runtime of the \textit{OmpSs-2} task-based programming model to integrate our Linux Kernel extension. The original Nanos6 threading model relies on explicit management of thread binding. The runtime keeps a single worker bound to each core and an unbounded Leader Thread that periodically wakes up to run polling services\cite{bsc:tampi}. Workers continuously request tasks and only leave the core voluntarily when: no more tasks are left, an explicit {\it taskwait} construct prevents the task to continue until all its children tasks complete, or the next ready task to execute is already bound to another worker (in which case there is a swap of workers). The Nanos6 threading model has been extended to manage multiple workers bound to the same core in order to support UMT.

The UMT-enabled Nanos6 initialization phase still creates a worker bound at each core and an unbounded Leader Thread, extended with the purpose of monitoring all eventfds. The Leader Thread main loop first performs a blocking read operation on all eventfds using the standard \textit{epoll} system call. When one of the monitored threads sends an event, the Leader Thread gets unblocked, reads the corresponding eventfd, subtracts the two eventfd counters, and calculates the number of ready threads by adding the result to the user-held counter. Then, for each core, it checks whether the count of ready threads is zero. If this is the case and there are still tasks to execute, the Leader Thread retrieves an idle thread from a pool and gives it a task to execute on the idle core. This is the precise point in which we break the original Nanos6 threading model where there could be a single ready thread per core at a time. In other words, in the case that a blocked worker was bound in the core where the Leader Thread has just started a second worker, it might happen that when the blocked worker resumes, the second one is still running and both of them have to compete for the core. However, in general, oversubscription prevails only for a limited amount of time.

Nanos6 workers perform the oversubscription check on every task scheduling point, which includes: starting, finishing or creating a task, as well as waiting (taskwait pragma) or yielding (taskyield pragma). To do so, workers first update the ready thread counters by doing a non-blocking read on the eventfds. Then, if the number of ready threads on the current core is greater than one, the worker self surrenders and returns to the pool of idle workers.

%Having multiple workers competing for the same core resources might pollute its cache and drop performance. However, assuming that the application defines multiple fine-grained tasks, the noise should not last much. In the Experimentation Section, we analysis the oversubscription noise and comment on the results.

\subsection{Discussion}
\label{sec:discussion}

The proposed design and implementation are based on several relaxed assumptions that simplify the design and do not particularly compromise performance. However, in future versions, we plan to improve them.

The core counters might occasionally suffer from temporal inconsistencies due to migrations or concurrent updates. For instance, if worker $A$ reads an eventfd of core 0, but gets preempted before it can update the shared Nanos6 atomic counter, another worker $B$ could use the current core counter value to determine whether to become idle or not. As a consequence, worker $B$ will take a decision based on incorrect data (the eventfd has been read, but the corresponding user-space core counter has not been updated yet). This could be solved by protecting this critical region with a lock, but we have declined this option given that the situation is unlikely to happen and non-critcal, i.e. there are two possible outcomes: A worker becomes idle when it is the only worker on its core, and a worker continues to run although its core is already being used. In the former case, the Leader thread would eventually notice the idle core during its 1ms periodic scans and schedule a worker there. In the latter case, the oversubscription period would simply last a bit longer. However, in any case, the application correctness would not be compromised.

With the objective of reducing cache pollution, it could be interesting to have a Leader Thread monitoring each core eventfd instead of having a single Leader Thread that monitors all of them. However, this would require much more Leader Thread context switches. For instance, in the single Leader Thread approach, if events have been generated in four different cores, a single Leader thread context switch will serve all of them. Instead, in the multiple Leader Thread approach, four context switches would be needed. Because a Leader thread context switch might lead to the preemption of a busy runtime worker, it is not clear whether having multiple Leader Threads would improve performance.

Another relevant technique that the multi-leader-thread and UMT-enabled Nanos6 might benefit from is the \textit{leader-follower} approach~\cite{schmidt2000}. Sometimes, the Leader Thread might attempt to wake up a worker in the same core that it is running. Instead, the Leader Thread could first create or designate another worker to become the new Leader and then, it could morph into a standard worker to immediately start executing tasks. The recently nominated Leader Thread would wake up at some point to continue listening for incoming UMT events, repeating the cycle again\footnote{This is quite similar to how SA proposes to respond to kernel events.}.

The proposed design notifies user-space whenever a worker blocks or unblocks. However, Nanos6 workers only need to take immediate action when the counter reaches zero. Subsequently, it would be interesting to adapt the Linux kernel to notify user-space only when there are no ready workers bound to a core. This mechanism would also outdate the mentioned eventfd overflow issue.

Please note that retaining worker within the kernel when they are unlocked (just as it is done in SA) to prevent oversubscription would enable the possibility of deadlocking the application and the system as commented in Section~\ref{sec:related-works}. Therefore, UMT avoids holding workers within the kernel; it just makes more information available to user-space.

\noindent In summary, the main UMT advantages are:
\begin{itemize}
\item Simple and lightweight extension to monitor blocking and unblocking events of threads.
\item Deadlock free, compared to similar methods such as SA.
\end{itemize}

\noindent And the disadvantages:
\begin{itemize}
    \item Periods of oversubscription might impact performance depending on the application, but simple techniques can keep it to a minimum (see Section \ref{sec:optimi}).
    \item Unnecessary context switches, which can be solved with a leader-follower approach.
    \item Unnecessary block/unblock events, although this can be palliated by notifying only when the cores become idle.
    \item User-space core counter temporal inconsistencies, which, as explained above, are not critical and unlikely.
\end{itemize}

% Impelementation
% Explain the simple model using the eventfd around the linux scheduler. explain
% that we could have modified the clone syscall but was not necessary
% reapeat that is simple and small overhead, comment the preempt enable and
% disable needs

%%% Local Variables:
%%% mode: latex
%%% TeX-master: "sample-sigconf"
%%% End:

\section{Experimental validation}
\label{sec:experiments}

We evaluate the UMT performance on both Network and Storage I/O with two applications: The Full-Waveform Inversion (FWI) application mock-up and the Heat Diffusion benchmark based on the Gauss-Seidel algorithm.

Both applications use synchronous network operations instead of asynchronous and, therefore, must enforce sequential ordering of communications. Using efficiently asynchronous APIs with task-based programming models requires complex code or special support\cite{bsc:tampi, bsc:tasio} that might not be available in all runtimes. Nonetheless, we show that UMT offers a generic approach to alleviate this problem transparently.

Network I/O over Omni-Path and Infiniband is directly managed in user-space, so it never blocks on the kernel side. Therefore, we have run all tests on top of an Ethernet network to illustrate the effect of UMT over network communications.

Performance metrics are obtained for all applications with both a UMT-enabled Nanos6 runtime and an unmodified version. Both runtimes are executed on top of the modified kernel, but only the former activates the kernel UMT facility.

\subsection{Environment, Tools and Metrics}
\label{sec:env}

All tests have been run on the BSC's "Cobi" Intel Scalable System Framework (SSF) cluster. Each node features two Intel Xeon E5-2690v4 sockets with a total of 28 real cores and 56 hardware threads at 2.60GHz, 128GiB of DDRAM 4 memory at 2400 MHZ, a 960 GB SSD Intel Optane 905P used to run the benchmarks, and an Intel DC S3520 SATA SSD with 222GiB that holds the system installation. All nodes are connected with both Intel Omni-path and Ethernet networks. The node's Linux distribution is a minimal installation of a SUSE Linux Enterprise Server (SLES) 12.2-0. We have updated the genuine distribution's kernel with our UMT-enabled Linux Kernel 5.1.

We profiled both the Optane and SATA SSDs maximum random read and write speeds using the Flexible I/O tester (fio)~\cite{bench:fio} by running 56 threads issuing up to 4 asynchronous I/O operations of 1MiB each. The Optane SSD approximately reported Reads of 2500 MiB/s and write of 2100 MiB/s. The SATA SSD showed an approximate peak performance of 250 MiB/s for writes and 270 MiB/s for reads.

Custom metrics (such as oversubscription period) were obtained with the Linux Trace Toolkit next generation (LTTng)~\cite{desnoyers:lttng} and the Babeltrace parser. The visualization tool Trace Compass~\cite{web:tc} was a key component to analyze UMT.

\subsection{Oversubscription and optional tweaks}
\label{sec:optimi}
UMT works transparently without any application modifications. However, minor straightforward programming techniques will increase its performance by limiting oversubscription and enabling more parallelism.

The application developer only needs to consider the following task layout to keep \textit{oversubscription} to a minimum: Avoid packing I/O operations and computationally intensive work, in this order, within the same task. Instead, it is better to either split I/O and computation into two different tasks or, if possible, do computation first and then I/O. Another option is to add a task scheduling point after the I/O such as a taskyield or a taskwait (that, in general, will not impact performance as it will not wait for any task).

The reasoning is twofold: On the one hand, blocking operations trigger the UMT mechanism resulting in additional awakened workers running more tasks. On the other hand, the UMT-enabled Nanos6 oversubscription prevention mechanism only forces workers to surrender at task scheduling points (such as task finish). Therefore, tasks that perform I/O followed by computation will first trigger the execution of multiple threads per core that will immediately block while performing I/O, triggering the wake up of further workers until either no more runnable tasks or workers are left. Then, eventually, threads will gradually wake up as their requested data becomes available and will resume execution. If the second part of the tasks is computationally intensive, all the previously waken up threads will have to inevitably compete for a share of their assigned core. The oversubscription period will continue until the first tasks finish and Nanos6 has a chance to stop workers before they get another task. Alternatively, adding a scheduling point between I/O and computation will enforce an oversubscription check within the tasks execution, which will also prevent the problem.

Task-based MPI applications usually need to enforce \textit{sequential ordering on their communication tasks} as it is possible that all cores assigned to two MPI processes became blocked while running unmatched MPI send and receive operations. When UMT is in use, there is no need for such restriction as long as networking transmissions do block. In such cases, UMT will report idle cores due to either blocking send or receive operations, and the runtime will be able to schedule more tasks. Eventually, all matching send-receive operations will be in-flight, and the execution will continue.

UMT seamlessly overlaps I/O and computation, as long as there is enough parallelism. To enable more parallelism (if needed) implementing a \textit{n-buffering scheme} might be useful in order to defer I/O operations while preventing stalls on dependent task. This is particularly useful for write operations, as read operations are likely to be in the critical path.

\subsection{UMT and the page cache}
\label{sec:pc}
UMT also works with I/O indirectly performed by the page cache. When a worker writes data to the page cache, and it is full, the thread blocks until there is enough free memory to proceed. Because UMT reports blocking threads regardless of the cause, the runtime is notified on this situation and it responds as usual. However, the page cache flush might be performed by a kernel thread when the amount of system memory is below a certain threshold, or when pages are older than a configurable number of centiseconds\footnote{See dirty\_background\_ratio, dirty\_ratio, dirty\_expire\_centisecs and dirty\_writeback\_centisecs in the Linux Kernel source Documentation/sysctl/vm.txt file}. In such cases, runtime threads do not block because flushing is performed transparently by the system which, in fact, is also overlapping I/O with computation. Yet this approach, unlike UMT combined with non-buffered I/O, does not extend to read operations and has the additional cost of an extra memory copy for writing to the page cache. An advantage of the page cache is that it naturally optimizes write I/O operations whose address coincide in the same page cache before they are flushed, reducing the total number of bytes written to the storage device. However, this only applies to applications that write multiple times to the same addresses. We further evaluate the effect of UMT and the page cache in Section~\ref{sec:res-gauss-intro}.

%%%%%%%%%%%%%%%%%%%%%%%%%%%%%%%%%%%%%%%%%%%%%%%%%%%%%%%%%%%%%%%%%%%%%%%%%%%%%%%
%%%%%%%%%%%%%%%%%%%%%%%%%%%%%%%%%%%%%%%%%%%%%%%%%%%%%%%%%%%%%%%%%%%%%%%%%%%%%%%
%%                         ________        _____
%%                         |  ___\ \      / /_ _|                          
%%                         | |_   \ \ /\ / / | | 
%%                         |  _|   \ V  V /  | | 
%%                         |_|      \_/\_/  |___|
%%
%%%%%%%%%%%%%%%%%%%%%%%%%%%%%%%%%%%%%%%%%%%%%%%%%%%%%%%%%%%%%%%%%%%%%%%%%%%%%%%
%%%%%%%%%%%%%%%%%%%%%%%%%%%%%%%%%%%%%%%%%%%%%%%%%%%%%%%%%%%%%%%%%%%%%%%%%%%%%%%

\subsection{Full Waveform Inversion Mock-up (FWI)}

\subsubsection{Introduction}
\label{sec:fwi-intro}

The acoustic Full Waveform Inversion (FWI) \cite{fwi} method aims to generate high-resolution subsoil velocity models from acquired seismic data through an iterative process. The time-dependent seismic wave equation is solved forward and backward in time in order to estimate the velocity model. %From the differences between acquired data and the computed velocity model, a gradient volume is calculated and used to update the velocity model on the next iteration.

%The inverse problem is nonlinear and ill-conditioned. This makes it difficult solving the problem
%at high frequencies. Instead, the initial stimulus is decomposed into a spectrum of frequencies.
%Then, low frequencies are solved first on a coarse grid providing a good guess for higher frequencies.

%Conceptually, FWI can be divided into three main steps. A pre-processing step estimates the 
%computational resources needed to solve the problem according to the number of shots, wavelet
%frequency and domain dimensions. Then, the wave propagator solves the time domain formulation
%of the wave equation forward and backward in time. Finally, a post-processing step gathers the
%information from the computation of all different frequencies into a single final velocity model.
%The workflow is shown in Fig. \ref{fig:fwi_over}.
%
%All three stages of the FWI require intensive I/O operations. While pre and post-processing steps
%perform sequential read and write operations on large shared velocity model files, the wave
%propagator mostly performs local I/O. From the computational point of view, the pre and
%post-processing stages do not represent a major issue on the performance of the FWI.

From the computational point of view, FWI is mainly divided into two phases: the forward propagation and backward propagation. On each phase, two three-dimensional volumes are updated for a sequence of time steps, one for velocities and one for stresses. In each forward propagation timestep, the FWI models are updated and an snapshot might be saved to disk. Next, in the backward propagation phase, all timesteps are processed again but in inverse order and corresponding snapshots are read instead of being written.

% ==== removed because of lack of space
%More specifically, a timestep is composed of four steps. First the velocity volume is written/read to/from disk. Then, it is updated with the stress model of the previous iteration. Next, the stress model is updated with the new velocity volume and, finally, the source wave is inserted into the stress volume.

% ======= Simplified FWI ompss-2 only description: ======
%We have parallelized the FWI mock-up with the OmpSs-2 task-based programming model. The computation of velocity and stress volumes have been divided into tasks at the y-plane level (slice). Snapshots are read and written in tasks that also work at the same slice level. Fig. \ref{fig:fwi_task} shows a timestep task decomposition for a forward propagation. The backward propagation task decomposition is analogous to the forward propagation and it is not shown. Notice that the figure shows the wave source insertion task, which is unique per timestep. It can be seen that the stress tasks can be computed in parallel with write tasks. However, velocity tasks depend on the write tasks because the velocity model data must be written to disk before it is updated with the next timestep iteration.

\begin{figure*}
\centering
\includegraphics[width=\textwidth]{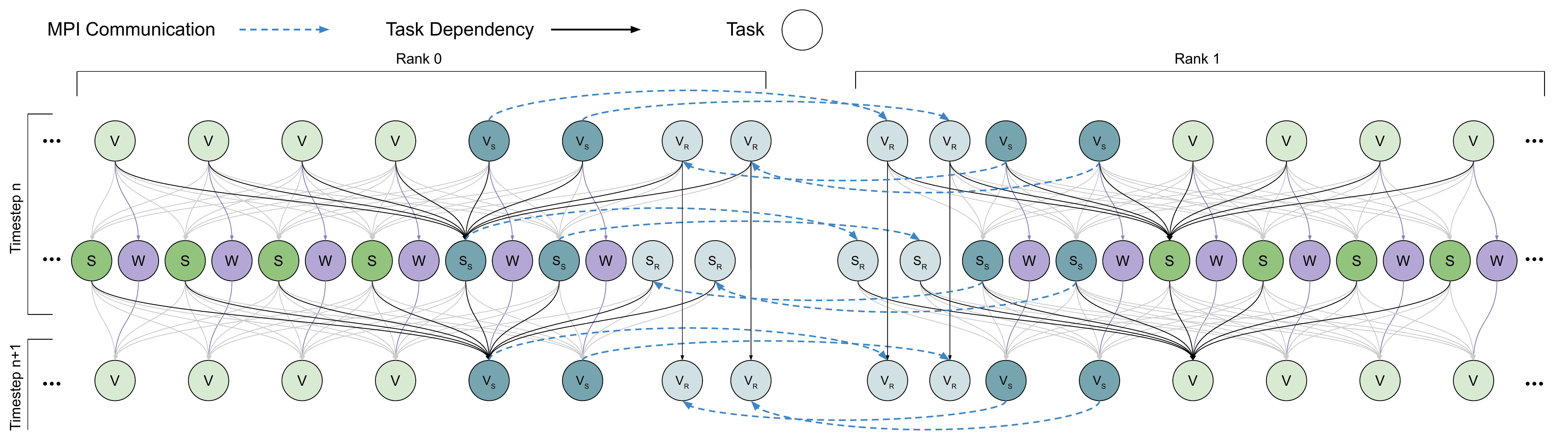}
% simplified caption due to lack of space
%\caption{FWI task and MPI decomposition. This figure shows the complete stencil used in the Full-Waveform Inversion algorithm and the MPI communications between two ranks. Each circle is a task. Tasks work exclusively with a single volume slice. \texttt{V} stands for a task enclosing the computation of a velocity slice. \texttt{S} is the equivalent task for the stress volume. \texttt{$V_{S}$} and \texttt{$S_{S}$} tasks perform the computation of a slice just as \texttt{V} and \texttt{S} do but they additionally send the slice afterwards through MPI to the required rank. \texttt{$V_{R}$} and \texttt{$S_{R}$} tasks' only purpose is to receive a slice sent through MPI. \texttt{W} task writes a velocity slice into disk. The task performing the source insertion in the stress volume is omitted for clarity. Task dependencies are shown as arrows. Dashed arrows show MPI communications. The dependencies for a single stencil step are colored in black for clarity, gray dependencies belong to other stencil steps of the same timestep. Purple arrows describe I/O dependencies.}

\caption{FWI task and MPI task decomposition of two ranks. Tasks work with a single volume slice. \texttt{V} and \texttt{S} compute velocity and stress slices respectively. The S suffixes denote tasks that compute and send its block with MPI. The R suffixes only receive a data block. \texttt{W} tasks write a velocity slice into disk.}
\label{fig:fwi_tasks}
\end{figure*}

We have parallelized the FWI mock-up using MPI and OmpSs-2. The velocity and stress volumes are split at the Y-slice level, being a slice of length 1 the minimum parallel granularity. Fig. \ref{fig:fwi_tasks} shows a timestep task decomposition for a forward propagation. The backward propagation is analogous and it is not shown. Each MPI rank is assigned a range of consecutive Y-slices for both the velocity and the stress volumes. After a rank has finished computing either the velocity or the stress slices, it sends the left-most and right-most slices to the previous and next ranks respectively. The exact number of exchanged slices in the halo is an execution parameter. All slices assigned to an MPI rank are computed in OmpSs-2 tasks. Each task wraps the computation of exactly one slice. Writing and reading snapshots is also done at the slice level. The velocity and stress tasks involved in the computation of the halo also perform the MPI send operation to the appropriate rank. Instead, the MPI receive tasks are run on standalone tasks.

%A single task per timestep performs the source insertion operation. However, it is not shown in the figure for simplicity. It can be seen that the stress tasks \texttt{S} can be computed in parallel with write (or read) tasks \texttt{W}. However, velocity tasks \texttt{V} depend on the write tasks because the velocity model data must be written to disk before it is updated with the next timestep iteration. As we explain later, we have used buffering to alleviate this. Also, MPI receive tasks \texttt{$V_{R}$} and \texttt{$S_{R}$} depend on its homologous of the next timestep (the dependency is only shown for the velocity tasks) to avoid multiple receive tasks operating on the same slice to be running at the same time.

%The advantage of working at the slice level, even for MPI communications and I/O operations is to avoid the usage of barriers that would break the natural flow of dependencies. Taskification allows to start the computation of slices at timestep $i+1$ as soon as all the data it depends on is ready, even if not all tasks of timestep $i$ have completed. With task based programming models this is as simple as specifying the data used for input and/or output of each tasks and let the runtime determine the flow of tasks necessary to complete the execution.

We have introduced the three optimizations targeting UMT mentioned in Section \ref{sec:optimi}.

\subsubsection{Results}

We have evaluated FWI on two scenarios: A single node which involves only storage I/O and two nodes, which also includes network I/O. In both scenarios, we use one MPI process per socket to maximize the main memory bandwidth and restrict each MPI processes to work with its own file. On each scenario, tests are repeated for both SATA and Optane SSDs. In all cases, performance is evaluated in terms of processed kilo volume cells per second (kc/s). We have run between 5 and 10 repetitions for each test.

We used two problem sizes; one for the single and another for the two-node settings, being the later twice in size than the former. The input frequency is 20Hz for both cases and the volume dimensions in terms of Z, X and Y are 208x208x408 for the single node tests and 208x208x808 for the two-node test. Each I/O task processes a Y-slice of 1521 KiB being the total volume size approximately 606 MiB large. In total, 118 forward and 118 backward propagation iterations are processed. All tests are run with an I/O frequency (iof) of 1 and 3. In total, each node writes and reads 70GiB for 1 iof and 23GiB for 3 iof.

%todo: \hl{ get oversubscription ratio}

% simplified becacuse of lack of space
%It is discouraged to use the OS page cache with the FWI I/O operations. Although FWI rereads in the backward phase what was written in the forward phase, it is likely that all the cached data is flushed and replaced by the time it is attempted to be reused. Typically, operating systems implement variants of the Last Recently Used (LRU) algorithm in which the evicted pages are the last used ones. However, for big problem sizes, the entire page cache might be flushed entirely several times by the time the backward phase starts. Although it is true that the last iterations of the forward phase and the first ones of the backward phase will be able to reuse some data, the improvement is likely to be negligible considering the page cache overhead. We further evaluate the effect of UMT and the page cache in Section~\ref{sec:res-gauss-intro}

The use of the page cache is discouraged on the FWI. Although FWI rereads in the backward phase what was written in the forward phase, it is likely that all the cached data is flushed and replaced by the time it is attempted to be reused.

%We run a preliminary series of experiments with and without the page cache and proved that the use of non-buffered I/O is either equal or slightly better than buffered.

Table \ref{tab:fwi-results} summarizes the results obtained on both SATA and Optane SSDs. UMT improves the performance on almost all the presented scenarios. The key factor of all those metrics is that the baseline's core usage is suboptimal even though the task-based parallelization approach and task granularity should be enough to keep all cores busy during the entire execution. In consequence, it can be deduced that idle time is spent on I/O operations. The UMT-enabled Nanos6 runtime is aware of idle cores and uses such knowledge to schedule pending work on them, effectively improving resource usage, which in most cases almost reaches 100\%.

\begin{table*}[]
\caption{FWI results. Disk throughput is reported per node. Disk and Network are reported in MiB/s.}
\begin{tabular}{|c|c|c|c|c|c|c|c|c|c|c|c|c|c|}
\hline
\multirow{3}{*}{\textbf{Storage}} & \multirow{3}{*}{\textbf{IOF}}  & \multirow{3}{*}{\textbf{Version}} & \multicolumn{5}{c|}{\textbf{One Node}}                                                                                                                   & \multicolumn{6}{c|}{\textbf{Two Nodes}}                                                                                                                                                          \\ \cline{4-14} 
				  &                                &                                   & \multicolumn{2}{c|}{\textbf{FOM}}                                            & \multirow{2}{*}{\textbf{CPU(\%)}} & \multicolumn{2}{c|}{\textbf{Storage I/O}} & \multicolumn{2}{c|}{\textbf{FOM}}                                            & \multirow{2}{*}{\textbf{CPU(\%)}} & \multicolumn{2}{c|}{\textbf{Storage I/O}} & \multirow{2}{*}{\textbf{Network I/O}} \\ \cline{4-5} \cline{7-10} \cline{12-13}
                                  &                                &                                   & \multicolumn{1}{l|}{\textbf{Speedup}} & \multicolumn{1}{l|}{\textbf{kc/s}}   &                                   & \textbf{w}          & \textbf{r}          & \multicolumn{1}{l|}{\textbf{Speedup}} & \multicolumn{1}{l|}{\textbf{kc/s}}   &                                   & \textbf{w}          & \textbf{r}          &                                       \\ \hline
\multirow{4}{*}{\textbf{Optane}}  & \multirow{2}{*}{\textbf{3}}    & \textbf{Baseline}                 & \multirow{2}{*}{13\%}                 & 12103                                & 74.66                             & 141.14              & 140.57              & \multirow{2}{*}{38\%}                 & 19407                                & 51.01                             & 112.10              & 113.77              & 45.04                                 \\ \cline{3-3} \cline{5-8} \cline{10-14} 
                                  &                                & \textbf{UMT}                      &                                       & 13714                                & 96.64                             & 157.08              & 162.25              &                                       & 26677                                & 93.07                             & 154.69              & 155.79              & 61.80                                 \\ \cline{2-14} 
                                  & \multirow{2}{*}{\textbf{1}}    & \textbf{Baseline}                 & \multirow{2}{*}{13\%}                 & 11883                                & 75.01                             & 409.21              & 406.77              & \multirow{2}{*}{34\%}                 & 19492                                & 52.57                             & 328.29              & 341.17              & 45.21                                 \\ \cline{3-3} \cline{5-8} \cline{10-14} 
                                  &                                & \textbf{UMT}                      &                                       & 13399                                & 97.22                             & 457.21              & 462.25              &                                       & 26165                                & 94.08                             & 453.09              & 445.38              & 60.58                                 \\ \hline
\multirow{4}{*}{\textbf{SATA}}    & \multirow{2}{*}{\textbf{3}}    & \textbf{Baseline}                 & \multirow{2}{*}{15\%}                 & 11782                                & 74.98                             & 138.24              & 136.01              & \multirow{2}{*}{39\%}                 & 19157                                & 51.90                             & 110.96              & 112.00              & 44.46                                 \\ \cline{3-3} \cline{5-8} \cline{10-14} 
                                  &                                & \textbf{UMT}                      &                                       & 13506                                & 96.35                             & 154.76              & 159.64              &                                       & 26582                                & 93.69                             & 154.59              & 154.78              & 61.60                                 \\ \cline{2-14} 
                                  & \multirow{2}{*}{\textbf{1}}    & \textbf{Baseline}                 & \multirow{2}{*}{1\%}                  & 7450                                 & 55.28                             & 249.74              & 262.17              & \multirow{2}{*}{4\%}                  & 14240                                & 52.28                             & 233.03              & 257.23              & 33.04                                 \\ \cline{3-3} \cline{5-8} \cline{10-14} 
                                  &                                & \textbf{UMT}                      &                                       & 7561                                 & 70.55                             & 252.33              & 267.33              &                                       & 14752                                & 68.42                             & 243.93              & 263.72              & 34.20                                 \\ \hline
\end{tabular}
\label{tab:fwi-results}
\end{table*}

Because work is computed earlier in UMT, more pressure falls on I/O devices. On systems where such devices are not saturated, both network and storage devices can increase its throughput resulting in a generalized improved performance. However, as can be seen by looking at the SATA test with 1 iof, no further improvements are achieved if the storage device is already at its peak performance on the baseline version. Indeed, that I/O can still be overlapped with computation but, if the storage device is the bottleneck, the UMT effect is limited to compute work in advance, although it will still be needed to wait for the storage device in the end.

The two-node setting achieves particularly high speedups due to two main reasons: On the one hand, Ethernet communications are likely to block and UMT overlaps other work in the meantime. On the other hand, the use of UMT disposes the need for network serialization (as explained in Section~\ref{sec:optimi}) which reduces the number of synchronization points.

\begin{figure*}[htb]
\centering
\includegraphics[width=\textwidth]{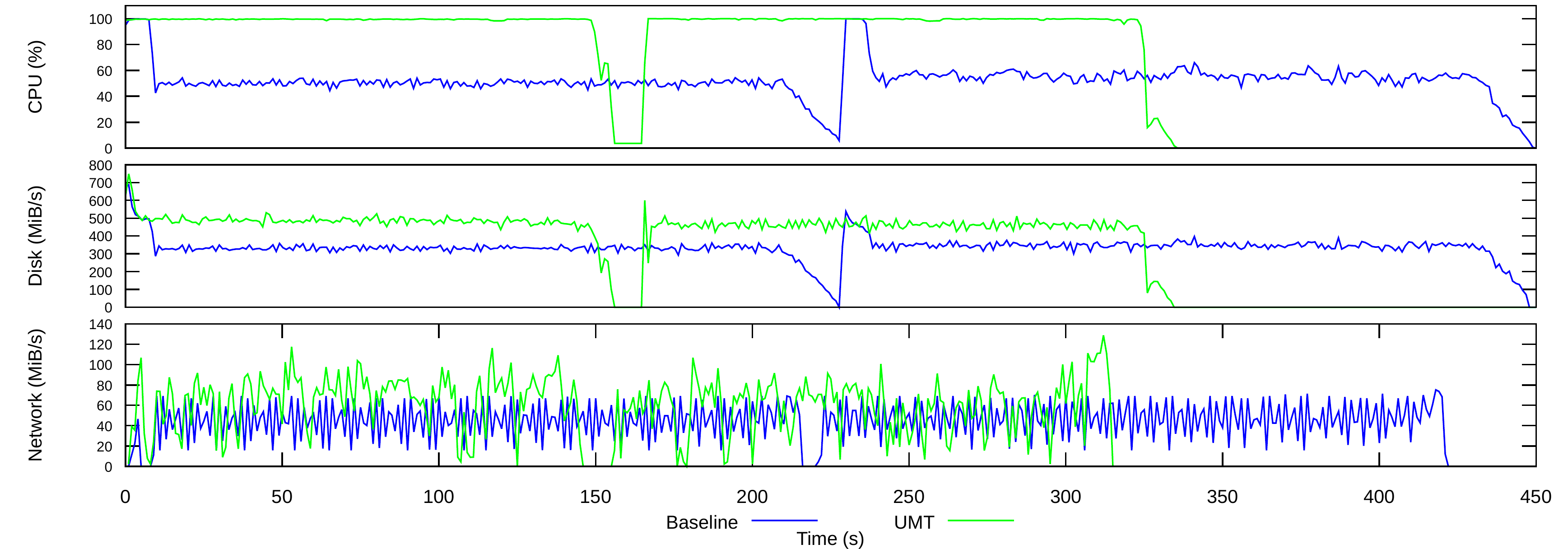}
\caption{FWI baseline vs UMT metrics run on two nodes backed by Optane.}
\label{fig:fwi-metrics}
\end{figure*}

Fig. \ref{fig:fwi-metrics} shows the FWI CPU, Disk and Network utilization graphs with and without UMT for a single run on two-nodes with Optane. The disk view aggregates both read and write operations, but the kind of I/O can be easily distinguished by the division in the graphs that separates both FWI phases: First, write for the forward phase and, second, read for the backward phase. The CPU view shows how UMT affects the execution of each phase by bringing its utilization to almost full capacity. CPU usage drops a bit at the end of each phase when the number of remaining tasks becomes low and it is not possible to continue overlapping I/O with computation. The network and disk views, far from being saturated, follow the same tendency and see its throughput increased during both phases.

Because UMT does not distinguish the block/unblock reason that triggers it, it is not possible to know to which extent UMT influences each I/O interface and how each of them individually contributes to the overall performance improvement. Also, their effects get combined; for instance, increasing CPU usage due to overlapping storage I/O with computation implies that work is computed earlier and, therefore, more data is ready to be served through the network I/O interface. In consequence, it is complex to dissect each component. However, with the purpose of focusing on storage I/O (instead of network I/O and the non-sequential ordering constraint effect) we have run two totally independent FWI instances on a single node (one per socket) with half of the problem size and iof 3 for SATA and iof 1 for Optane. The tests run on Optane showed 3\% speedup, while SATA tests obtained 6\%. Hence, it can be deduced that a considerable benefit is obtained from communications.

Oversubscription impact has been negligible for all tests. Our custom LTTng and Babeltrace analysis scripts reported UMT oversubscription periods limited to approximately 2.25\% of the total execution length. The number of per core context switches was incremented by 8200 in 330s, approximately.

Finally, the Linux Perf tool has been used to analysis the FWI, Nanos6 and Linux Kernel individual performance. Table~\ref{tab:perf_over} shows three Perf traces with sampling frequencies grouped by their dynamic shared object. We used multiple frequencies because although high sampling frequency increases the trace accuracy, it also generates bigger traces that might affect the storage device performance. The table shows that the UMT-enabled Nanos6 and Linux Kernel overhead increase is just slightly higher than the baseline versions, approximately a 0.04\% for Nanos6 and 0.10\% for the Linux Kernel.

\begin{table}
\caption{Percentage of Perf samples distributed over FWI, Nanos6 and Linux Kernel for three sampling frequencies.}
\centering
\begin{tabular}{|r|c|c|c|}
\hline
Component		& 99Hz(\%) & 999Hz(\%) & 9999Hz(\%) \\  \hline
\hline
FWI (baseline)		& 93.58	& 97.05	& 97.57	\\
FWI (UMT)		& 96.16	& 96.95	& 96.64	\\
\hline
Nanos6 (baseline)	& 0.05	& 0.06	& 0.06	\\
Nanos6 (UMT)		& 0,11	& 0.11	& 0.09	\\
\hline
Kernel (baseline)	& 0.46	& 0.39	& 0.40	\\
Kernel (UMT)		& 0.59	& 0.47	& 0.50	\\
\hline
\end{tabular}
\label{tab:perf_over}
\end{table}

\subsection{Heat diffusion}

\subsubsection{Introduction}
\label{sec:res-gauss-intro}

We use a Gauss-Seidel based iterative heat equation solver with checkpointing parallelized with OmpSs-2 and MPI following a wavefront strategy.

\begin{figure}[htb]
\centering
\includegraphics[width=\columnwidth]{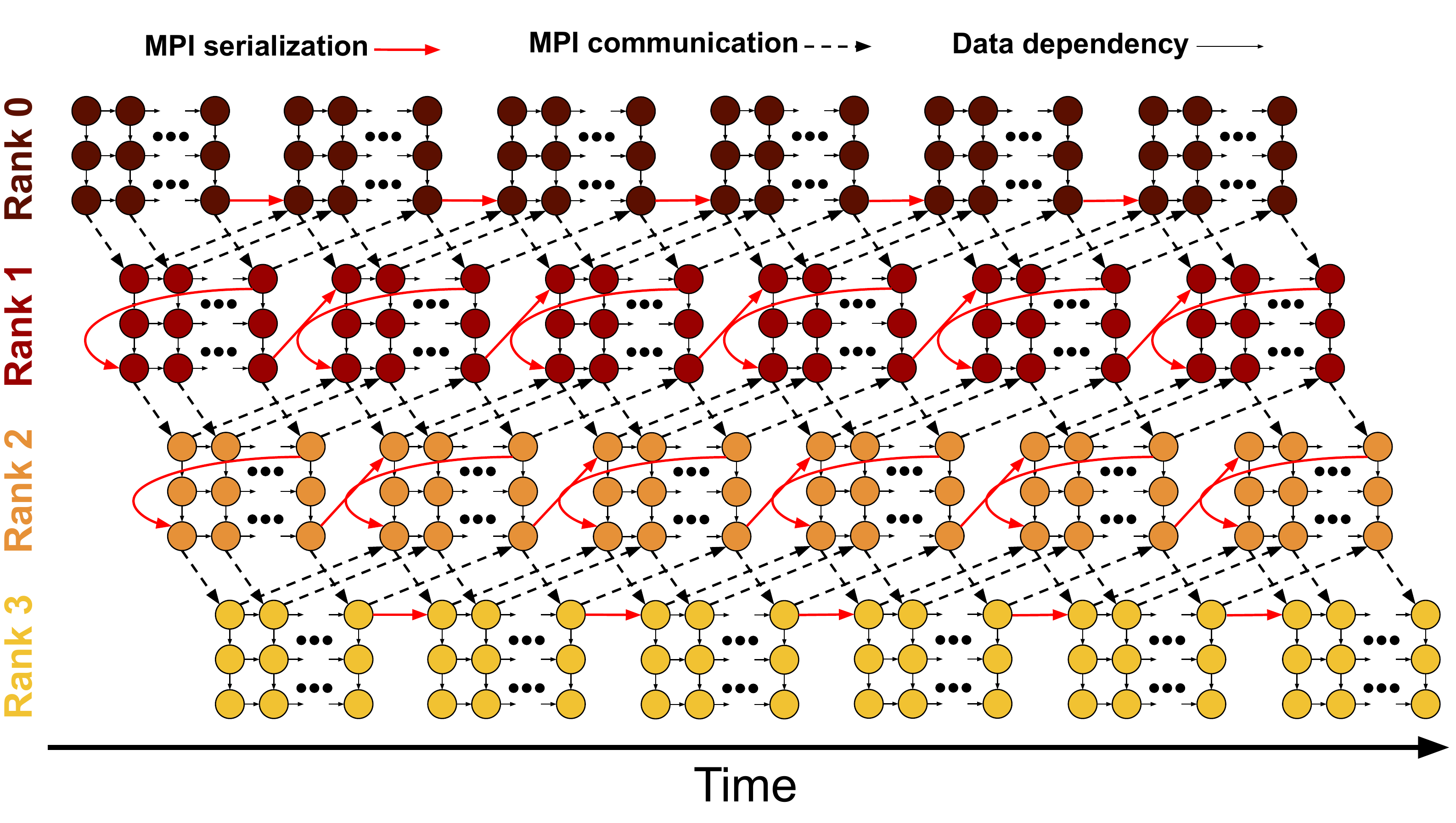}
\caption{Heat Diffusion task decomposition for four MPI processes. Tasks are shown as circles, there is no distinction between computation tasks and I/O tasks in terms of the task layout.}
\label{fig:gauss-tasks}
\end{figure}

Fig. \ref{fig:gauss-tasks} shows the task decomposition. Essentially, this algorithm updates a two-dimensional matrix for a number of iterations. Blocks of consecutive rows are distributed among MPI processes, who interchange halos with their previous and next rank. Each MPI process computes its share of the volume by splitting them into blocks that are processed in tasks. The execution of iterations is fully overlapped; the tasks that belong to the iteration $i+1$ start running as soon as the tasks it depends on from the iteration $i$ have been computed.

Checkpointing is performed every n-iterations by writing the whole model in parallel. When a checkpointing iteration is reached, a set of tasks that perform both the model update and storage I/O (in this order) are created instead of the usual update-only tasks. Because enough parallelism was available, there was no need to implement a n-buffering scheme.

This benchmark main purpose is to illustrate the UMT behaviour on applications that perform I/O mostly for checkpointing. This common I/O pattern is particularly interesting because write operations do not stall parallelism and easily enable the possibility of overlapping I/O with other tasks. 

Data written in each checkpointing iteration is usually not read again during the execution of the application and, therefore, the system's page cache is only adding the overhead of an extra memory copy (similarly as FWI). For this reason, this kind of write operations benefit from a non-buffered policy, but it also implies that I/O becomes a blocking operation which increases the time in which cores are idling. In consequence, checkpointing is, by its nature, an ideal scenario for UMT.

To illustrate such results, Table \ref{tab:gauss-buf} reports the Heat Diffusion performance obtained when running on top of the page cache (without O\_DIRECT). The remaining experiments on this section, rely on the Linux Kernel O\_DIRECT mechanism to bypass the Linux Kernel page cache. Please, note that although the buffered version achieves a similar speedup than the non-buffered version, the actual performance of the buffered version is worst (smaller) than the non-buffered.

%\footnote{In the Linux Kernel, when memory is below the threshold specified in /proc/sys/vm/dirty\_background\_ratio flushing starts, but subsequent writes to the page cache still do not block. When the number of free pages drops below /proc/sys/vm/dirty\_ratio, further writes do block until enough space is available.}
%\footnote{Pages older than /proc/sys/vm/dirty\_expire\_centisecs are written every /proc/sys/vm/dirty\_writeback\_centisecs centiseconds}

\begin{table}[]
\centering
\caption{Heat Diffusion buffered vs non-buffered storage I/O analysis on a single node with Optane and IOF 15. CPU is shown in percentage and storage I/O in MiB/s.}
\begin{tabular}{|c|c|c|c|c|c|}
\hline
\multirow{3}{*}{\textbf{Cache}}        & \multirow{3}{*}{\textbf{Version}} & \multicolumn{4}{c|}{\textbf{Buffered}}                                                         \\ \cline{3-6} 
                                       &                                   & \multicolumn{2}{c|}{\textbf{FOM}}       & \multirow{2}{*}{\textbf{CPU}} & \textbf{Storage I/O} \\ \cline{3-4} \cline{6-6} 
                                       &                                   & \textbf{Speedup}      & \textbf{cells/s} &                              & \textbf{Write}       \\ \hline
\multirow{2}{*}{\textbf{Buffered}}     & \textbf{Baseline}                 & \multirow{2}{*}{14\%} & 4835            & 78.11                         & 1242.85              \\ \cline{2-2} \cline{4-6} 
                                       & \textbf{UMT}                      &                       & 5492            & 87.35                         & 1408.89              \\ \hline
\multirow{2}{*}{\textbf{Non-Buffered}} & \textbf{Baseline}                 & \multirow{2}{*}{20\%} & 4934            & 76.36                         & 1239.09              \\ \cline{2-2} \cline{4-6} 
                                       & \textbf{UMT}                      &                       & 5949            & 90.79                         & 1485.96              \\ \hline
\end{tabular}
\label{tab:gauss-buf}
\end{table}

\subsubsection{Results}

Table \ref{tab:gauss-results} shows the results observed. Similarly to FWI, UMT improves the benchmark's performance by reducing the core's idle time while rising both storage and network I/O throughput.

We used two different input sets for the SATA and Optane SSDs. Because the Optane SSD is approximately 10 times faster than the SATA SSD, we had to increase the benchmark's checkpointing frequency while running with Optane to keep the pressure on the storage I/O; otherwise the benchmark's computational part would fairly extend the I/O part and the UMT's margin for improvement becomes too small for its effect to be appreciated. Optane tests write approximately 780GiB, while SATA tests write 117GiB, per node.

\begin{table*}[]
\centering
\caption{Heat Diffusion results. Disk throughput is reported per node. Disk and Network are reported in MiB/s.}
\begin{tabular}{|c|c|c|c|c|c|c|c|c|c|c|c|}
\hline
\multirow{3}{*}{\textbf{Storage}} & \multirow{3}{*}{\textbf{IOF}}   & \multirow{3}{*}{\textbf{Version}} & \multicolumn{4}{c|}{\textbf{One Node}}                                                     & \multicolumn{5}{c|}{\textbf{Two Nodes}}                                                                                            \\ \cline{4-12} 
				  &                                 &                                   & \multicolumn{2}{c|}{\textbf{FOM}}   & \multirow{2}{*}{\textbf{CPU(\%)}} & \textbf{Storage I/O} & \multicolumn{2}{c|}{\textbf{FOM}}   & \multirow{2}{*}{\textbf{CPU(\%)}} & \textbf{Storage I/O} & \multirow{2}{*}{\textbf{Network I/O}} \\ \cline{4-5} \cline{7-9} \cline{11-11}
                                  &                                 &                                   & \textbf{Speedup}  & \textbf{cells/s} &                               & \textbf{write}       & \textbf{Speedup}  & \textbf{Metric} &                               & \textbf{write}       &                                       \\ \hline
\multirow{2}{*}{\textbf{Optane}}  & \multirow{2}{*}{\textbf{15}}    & \textbf{Baseline}                 & \multirow{2}{*}{20\%} & 4934        & 76.36                         & 1239.09              & \multirow{2}{*}{97\%} & 5850        & 46.42                         & 739.85               & 0.64                                  \\ \cline{3-3} \cline{5-7} \cline{9-12} 
                                  &                                 & \textbf{UMT}                      &                       & 5949        & 90.79                         & 1485.96              &                       & 11522       & 87.62                         & 1438.67              & 0.64                                  \\ \hline
\multirow{2}{*}{\textbf{SATA}}    & \multirow{2}{*}{\textbf{100}}   & \textbf{Baseline}                 & \multirow{2}{*}{30\%} & 4017        & 63.23                         & 151.47               & \multirow{2}{*}{65\%} & 4531        & 37.12                         & 86.00                & 0.51                                  \\ \cline{3-3} \cline{5-7} \cline{9-12} 
                                  &                                 & \textbf{UMT}                      &                       & 5249        & 80.07                         & 196.83               &                       & 7496        & 59.46                         & 141.31               & 0.83                                  \\ \hline
\end{tabular}
\label{tab:gauss-results}
\end{table*}

\begin{figure*}[htb]
\centering
\includegraphics[width=\textwidth]{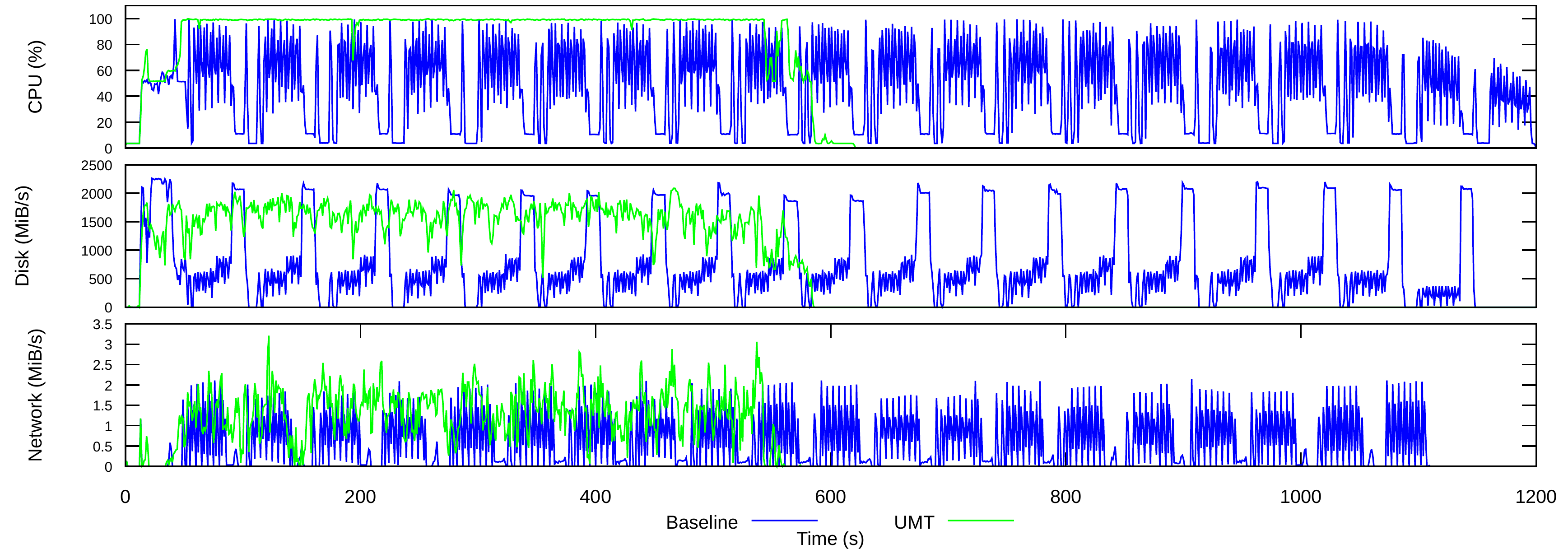}
\caption{Heat Diffusion baseline vs UMT metrics run on two nodes backed by Optane.}
\label{fig:gauss-metrics}
\end{figure*}

Fig. \ref{fig:gauss-metrics} shows the baseline vs UMT comparison of CPU, Disk and Network utilization for a full test run on two nodes with Optane. On the baseline version, CPU usage exhibits many bursts as the result of adjacent computing iterations. Checkpointing iterations dramatically increases idle time and, because they get interleaved with computation bursts, they prevent the cores from running at full capacity during most of the execution. The baseline Optane throughput follows the same pattern as the CPU usage, where its performance peaks coincide with CPU usage valleys. Again, a similar scenario repeats with the network. Instead, the UMT version flattens the CPU usage at its peak performance and keeps both Optane and Ethernet interface working at a higher capacity most of the time. As a result, performance almost reaches 2x (97\%).

The storage I/O only tests (two independent processes in one node with half the problem size each), reported 13\% speedup for SATA (iof 100) and 5\% speedup for Optane (iof 15). Again, showing that network communication improvements seem to be particularly relevant for these tests.

Similarly to FWI, oversubscription has had a negligible effect on performance. Oversubscription periods range between 2.4\% and 3.2\%. The number of per core context switches incremented by 115000 for two nodes (in 840s) and 35000 for one node (in 610s), approximately.

\section{Conclusion and perspectives}
\label{sec:conclusion}

In this paper, we presented the proof-of-concept of the User-Monitored Threads (UMT) model, which aims to monitor blocking and unblocking thread events based on the eventfd simplified pipe. The proposal introduces a simple and lightweight Linux Kernel extension along with a user-space runtime coupling. We implemented the kernel extension on top of a Linux Kernel 5.1, and adapted the OmpSs-2 task-based programming model runtime to make use of the Linux Kernel extension. The Nanos6 runtime uses the mechanism to identify idle cores. To do so, it uses the block and unblock UMT events to keep track of the number of ready workers bound to each system's core. When a core becomes idle because, for example, all of its bound workers are blocked while performing I/O operations, the runtime schedules additional workers on them. Multiple workers bound to the same core might lead to oversubscription periods, but the runtime minimizes the effect by forcing workers to self-surrender of its core when multiple threads in the ready state are detected to be bound on it.

We tested UMT with two applications, and conclude that it has two main effects: On the one hand, it provides a mechanism to queue more I/O operations which brings the application's I/O throughput closer to the storage device maximum rate. On the other hand, blocked processes no longer stall the core where they were bound, and useful computations can be run instead, effectively enabling transparent I/O and computation overlapping. Oversubscription periods might limit performance but, as studied in the Experimentation Section, simple implementation techniques dramatically limit its effect. Overall, we were able to achieve speedups of almost 2x with the evaluated tests than combine both storage and network I/O.

It is worth noting that applications mostly benefit from UMT when: they are not already saturating neither the system's cores nor the I/O devices, the application exhibits enough parallelism to completely overlap I/O with computation and, ideally (although not mandatory), non-bufferd I/O can be used.

Regarding our future work, we plan to improve and polish some UMT details. In the first place, we plan to carefully study a UMT version where notifications are only sent when a core is idle. Then, we will consider implementing a Nanos6 leader-follower approach to minimize the number of unnecessary context switches. And finally, we will continue testing UMT with further task-based I/O intensive real applications. Eventually, we will propose UMT for inclusion into the Linux Kernel mainline repository.

\section*{Acknowledgment}

This project is supported by the European Union's Horizon 2021 research and innovation programme under the grant agreement No 754304 (DEEP-EST), the Ministry of Economy of Spain through the Severo Ochoa Center of Excellence Program (SEV-2015-0493), by the Spanish Ministry of Science and Innovation (contract TIN2015-65316-P) and by the Generalitat de Catalunya (2017-SGR-1481). Also, the authors would like to acknowledge that the test environment (Cobi) was ceded by Intel Corporation in the frame of the BSC - Intel collaboration.

\bibliographystyle{IEEEtran}
\bibliography{IEEEabrv,sigproc}

\begin{thebibliography}{10}
\providecommand{\url}[1]{#1}
\csname url@samestyle\endcsname
\providecommand{\newblock}{\relax}
\providecommand{\bibinfo}[2]{#2}
\providecommand{\BIBentrySTDinterwordspacing}{\spaceskip=0pt\relax}
\providecommand{\BIBentryALTinterwordstretchfactor}{4}
\providecommand{\BIBentryALTinterwordspacing}{\spaceskip=\fontdimen2\font plus
\BIBentryALTinterwordstretchfactor\fontdimen3\font minus
  \fontdimen4\font\relax}
\providecommand{\BIBforeignlanguage}[2]{{%
\expandafter\ifx\csname l@#1\endcsname\relax
\typeout{** WARNING: IEEEtran.bst: No hyphenation pattern has been}%
\typeout{** loaded for the language `#1'. Using the pattern for}%
\typeout{** the default language instead.}%
\else
\language=\csname l@#1\endcsname
\fi
#2}}
\providecommand{\BIBdecl}{\relax}
\BIBdecl

\bibitem{7967171}
J.~M. {Perez}, V.~{Beltran}, J.~{Labarta}, and E.~{Ayguadé}, ``Improving the
  integration of task nesting and dependencies in openmp,'' in \emph{2017 IEEE
  International Parallel and Distributed Processing Symposium (IPDPS)}, May
  2017, pp. 809--818.

\bibitem{anderson:SA}
T.~E. Anderson, B.~N. Bershad, E.~D. Lazowska, and H.~M. Levy, ``Scheduler
  activations: Effective kernel support for the user-level management of
  parallelism,'' \emph{ACM Transactions on Computer Systems (TOCS)}, vol.~10,
  no.~1, pp. 53--79, 1992.

\bibitem{williams:BSDSA1}
N.~J. Williams, ``An implementation of scheduler activations on the netbsd
  operating system.'' in \emph{USENIX Annual Technical Conference, FREENIX
  Track}, 2002, pp. 99--108.

\bibitem{Small:BSDSA2}
C.~A. Small and M.~I. Seltzer, ``Scheduler activations on bsd: Sharing thread
  management between kernel and application,'' 1995.

\bibitem{evans:FreeBSDSA}
J.~Evans and J.~Elischer, ``Kernel-scheduled entities for freebsd,'' 2000.

\bibitem{danjean:LinuxSA1}
V.~Danjean, R.~Namyst, and R.~D. Russell, ``Linux kernel activations to support
  multithreading,'' in \emph{In Proc. 18th IASTED International Conference on
  Applied Informatics (AI 2000}.\hskip 1em plus 0.5em minus 0.4em\relax
  Citeseer, 2000.

\bibitem{danjean:LinuxSA2}
------, ``Integrating kernel activations in a multithreaded runtime system on
  top of linux,'' in \emph{International Parallel and Distributed Processing
  Symposium}.\hskip 1em plus 0.5em minus 0.4em\relax Springer, 2000, pp.
  1160--1167.

\bibitem{danjean:LinuxSA3}
V.~Danjean and R.~Namyst, ``Controlling kernel scheduling from user space: An
  approach to enhancing applications’ reactivity to i/o events,'' in
  \emph{International Conference on High-Performance Computing}.\hskip 1em plus
  0.5em minus 0.4em\relax Springer, 2003, pp. 490--499.

\bibitem{WEB:LinuxSA}
\BIBentryALTinterwordspacing
B.~H. Ingo~Molnar. (2002) Linux kernel mailing list (lkml) discussion.
  [Online]. Available: \url{https://lkml.org/lkml/2002/9/24/305}
\BIBentrySTDinterwordspacing

\bibitem{rasiukevicius:BSDRemove}
M.~Rasiukevicius, ``Thread scheduling and related interfaces in netbsd 5.0,''
  2009.

\bibitem{windows:ums}
\BIBentryALTinterwordspacing
Microsoft. (2017) User-mode scheduling. [Online]. Available:
  \url{https://msdn.microsoft.com/en-us/library/windows/desktop/dd627187}
\BIBentrySTDinterwordspacing

\bibitem{appavoo:k42_sched}
J.~Appavoo, M.~Auslander, D.~DaSilva, D.~Edelsohn, O.~Krieger, M.~Ostrowski,
  B.~Rosenburg, R.~W. Wisniewski, and J.~Xenidis, ``Scheduling in k42,''
  \emph{White Paper, Aug}, 2002.

\bibitem{krieger:k42_build}
O.~Krieger, M.~Auslander, B.~Rosenburg, R.~W. Wisniewski, J.~Xenidis,
  D.~Da~Silva, M.~Ostrowski, J.~Appavoo, M.~Butrico, M.~Mergen \emph{et~al.},
  ``K42: building a complete operating system,'' \emph{ACM SIGOPS Operating
  Systems Review}, vol.~40, no.~4, pp. 133--145, 2006.

\bibitem{bsc:tampi}
K.~Sala, X.~Teruel, J.~M. Perez, A.~J. Pe{\~n}a, V.~Beltran, and J.~Labarta,
  ``Integrating blocking and non-blocking mpi primitives with task-based
  programming models,'' \emph{Parallel Computing}, vol.~85, pp. 153--166, 2019.

\bibitem{bsc:tasio}
A.~Roca, V.~Beltran, and S.~Mateo, ``Introducing the task-aware storage i/o
  (tasio) library,'' in \emph{International Workshop on OpenMP}.\hskip 1em plus
  0.5em minus 0.4em\relax Springer, 2019.

\bibitem{schmidt2000}
D.~C. Schmidt, C.~O'Ryan, M.~Kircher, I.~Pyarali, and B.~Back-endend,
  ``Leader/followers - a design pattern for efficient multi-threaded event
  demultiplexing and dispatching,'' in \emph{University of Washington.
  http://www.cs.wustl.edu/~schmidt/PDF/lf.pdf}.\hskip 1em plus 0.5em minus
  0.4em\relax AddisonWesley, 2000, pp. 00--29.

\bibitem{bench:fio}
J.~Axboe, ``Fio-flexible io tester,'' \emph{URL http://freecode.
  com/projects/fio}, 2014.

\bibitem{desnoyers:lttng}
M.~Desnoyers and M.~R. Dagenais, ``The lttng tracer: A low impact performance
  and behavior monitor for gnu/linux,'' in \emph{OLS (Ottawa Linux Symposium)},
  vol. 2006.\hskip 1em plus 0.5em minus 0.4em\relax Citeseer, 2006, pp.
  209--224.

\bibitem{web:tc}
\BIBentryALTinterwordspacing
T.~C. Community. (2019) Trace compass. [Online]. Available:
  \url{https://www.eclipse.org/tracecompass/}
\BIBentrySTDinterwordspacing

\bibitem{fwi}
J.~M. A.~Zitz, ``Final report on application experience in deliverable 6.3,
  deep-er project,'' 2017.

\end{thebibliography}

\end{document}